# Suppression of Supercontinuum Generation with Circularly Polarized Light

Arvinder S. Sandhu, Sudeep Banerjee and Debabrata Goswami*

*Tata Institute of Fundamental Research, Mumbai 400 005, India*

February 5, 2000

ABSTRACT

Controlling a nonlinear process like supercontinuum generation (SG) with the polarization-state of laser is an important demonstration of laser selectivity. We show that the threshold for SG and the total amount of supercontinuum generated depends on incident laser polarization for isotropic samples. Irrespective of the nature of the samples chosen, SG efficiency decreases as the incident laser polarization changes from linear to circular and thus, provides the first experimental demonstration of the suppression of SG with circularly polarized light. The ratio of the overall SG between the linear and circular polarization (i.e., measure of suppression) undergoes an intensity dependent decrease from large initial values to asymptotic limits, irrespective of samples.

*Phone: 022-215-2971, FAX: 022-215-2110, Email: debu@tifr.res.in

White light or supercontinuum generation (SG) [1-15] is a nonlinear phenomenon that occurs when the spectral content of a light pulse passing through a medium essentially "explodes" while remaining coherent at certain high intensities. SG using short laser pulses ($\leq 10^{-12}$ sec) has been extensively used in spectroscopy, ultrafast pulse generation, compression and amplification. Despite its widespread use, there are situations where SG is also considered a nuisance, such as in optical communications where it distorts channel signals [16]. Thus, it is important to control the process of SG. A lot of research has gone into understanding SG, and still a comprehensive theory is yet to emerge. Phenomenologically, in the condensed phase, SG can perhaps be understood as interplay of several processes, such as, self-phase modulation, four-wave mixing and multi photon excitation (MPE) etc. [1,2,5,7-11,13]. A recent study also evokes the band gap of the material as a necessary condition for SG [15].

This paper provides new insight into supercontinuum generation (SG) by femtosecond Ti:Sapphire laser pulses in condensed media. Our experimental results reveal a strong dependence of the SG on the polarization of the incident laser. We also provide detailed intensity dependence (over two orders of magnitude) for the SG process with a novel measurement technique. We find that SG is highly suppressed in the case of circular polarization as compared to the linear and the amount of suppression is intensity dependent. This is, to our knowledge, the first report on ellipticity dependence of the threshold and the efficiency of SG in isotropic samples. Laser polarization, thus, evolves as an important parameter for the control of the SG process. This result is of great relevance, considering the present trend of applications of SG, either in its more efficient generation or in suppression. In fact, suppression of continuum generation (or similar nonlinear processes) are especially important in the development of pulsed optical communications where the amount of intensity necessary for signal-transmission through useful fiber lengths is often enough to generate nonlinear distortions. Such nonlinear distortions happen to be one of the major bottlenecks towards the practical implementation of Terabit/sec data transmission capability that the pulsed optical communication technology can provide. Thus, the major focus of this paper is the demonstration of suppression and control on the amount of continuum generation (as a model nonlinear process) with incident laser polarization as a parameter, rather than the mechanistic study of continuum generation. Nevertheless, we do qualitative analysis of our observed results based on existing mechanisms for continuum generation. Polarization is just gaining more and more importance in the "control" community as an important parameter for controlling light-matter interaction, and the present work serves to be an important demonstration of its applicability for a highly nonlinear process like continuum generation.

We concentrate attention on two representative liquids, water and $CCl_4$, which have been extensively used for efficient SG [1,6,9,10,13]. The transparency of the material under study is important as it evades the question of damage and absorption of the incident and new wavelengths of light that are generated. We have also confirmed that the qualitative nature of most of our results presented here hold for optical fibers and other isotropic media like fused silica, $D_2O$, ethanol, etc. Our choice of $H_2O$ and $CCl_4$ as representatives for this paper are based on the fact that they are quite different from each other spectroscopically, and in properties like H-bonding, polarizability, band gap, etc. Consequently, the results presented here on polarization dependence of SG are independent of the choice of sample. All SG results reported in this paper pertain to measurements made using 100fs duration laser pulses. However, it is important to mention that we have verified that the gross polarization dependence mentioned in this paper also holds for 35ps laser pulses at 532nm.

We have used linearly polarized incident pulses from a chirped pulse amplified Ti:Sapphire laser (100fs, 806nm, 10Hz repetition rate) incident onto a 1-cm path length cuvette. A quarter-wave plate was used to vary the incident laser polarization from linear to circular. The laser peak powers (>100 MW) used in our experiments are above the self-focussing thresholds for both water and $CCl_4$. However, we have used large incident laser spot-sizes (2.5mm diameter at the sample) by placing the cuvette 7cm before the focal spot of the 50cm focussing lens (Figure 1). Consequently, we do not see SG to occur until the self-focussing process brings down the spot-size inside our sample to certain threshold intensities. The typical incident laser intensities at the input of the cuvette used in these experiments are in the range $7 \times 10^9 - 1.3 \times 10^{11}$ W/cm$^2$ (measured using a calibrated photodiode, PD1 in Figure 1). This intensity range was carefully chosen to reach the threshold for SG at the lower end, as well as to minimize heating, or other damage problems (e.g., dielectric breakdown of water $>10^{12}$ W/cm$^2$ [6]) at higher intensities. The large laser spot size has been chosen to make sure that the supercontinuum contribution is only from the sample and not from cuvette wall or air. We checked that the output polarization of the generated supercontinuum follows the input laser polarization for both the linear and circular cases, by using a broadband polarizer with an effective range of 400-700nm. Care has been taken to collect the entire white-light through imaging lenses either into a single-shot spectrometer (from Ocean-Optics Inc.) or onto a pulse photodiode (PD2 in Figure 1) with minimized residual fundamental. This was done by cutting-off most of the incident fundamental laser frequency either by spatial filtering or with the help of IR-filters (Figure 1). We did independent measurements without the sample for various laser parameters to check the robustness of the experimental setup for all spectral and PD measurements. It is found that the polarization response of the detectors and noise due to scattered light is within the error-limits of our data.

In Figure 2, we show two representative supercontinuum spectra of water for linearly and circularly polarized incident pulses at identical incident intensities. Each spectrum is an average over 20 laser shots. It is important to note the presence of Inverse Raman dip at ≈630 nm (corresponding to 3650cm$^{-1}$ O-H stretching mode) [6] in both of these spectra. The overall spectral content, or spread, of the SG for both polarizations is qualitatively similar; only the intensity of the generated wavelengths is polarization dependent. Thus, under these conditions, the ratio of the SG between the linearly and circularly polarized light is wavelength independent. It is, therefore, possible to make a good comparison between the linear and circular polarization cases by just monitoring the total SG as a function of the incident laser intensities instead of analyzing the individual wavelength components of the SG.

The overall SG was measured as a function of the incident laser intensity with the help of a dual PD arrangement (Figure 1). In this scheme, one of the PDs monitors the intensity of the incident laser while the other simultaneously measures the total SG for every laser shot and a software developed in our lab is used to correlate every input pulse to the output continuum pulse. This dual PD data-acquisition scheme enables us to take into account the pulse to pulse intensity fluctuations. Figure 3a shows the plot of total supercontinuum generated ($I_{SG}$) with respect to the incident laser intensity ($I_p$) for water and $CCl_4$. It is seen from figure 3a that the threshold for SG in $CCl_4$ is much lower than that of water and at any given $I_p$, the $I_{SG}$ for $CCl_4$ is at least an order of magnitude higher.

It is also important to note that the threshold for SG using circularly polarized light is always higher than that obtained using linearly polarized light. Thus, we can think of laser polarization as an on-off switch for the SG process, if our input intensities are in between the threshold levels for the linear and circular polarization.

As discussed earlier, SG process is theoretically understood by different authors as MPE enhanced SPM process [13], four wave mixing processes [10] or avalanche ionization enhanced SPM [14]. However, our experimental collection procedure minimizes the contribution from the wavelengths near fundamental, which is mostly characterized by SPM. Additionally, as intensities increase, SG occurs in a conical rainbow like emission, suggesting four-wave mixing as an important process in our present setup. The presence of Inverse Raman dip at ≈630 nm (Figure 2) also suggests the presence of Raman activity at high intensities. Thus, in order to explain the above experimental observations qualitatively, let us invoke the nonlinear susceptibility constants that have been used extensively for modeling SG [5,10] based on the four wave mixing. In an isotropic medium, the first nonlinear correction to the refractive index ($\Delta n$) comes from $\chi^{(3)}$, the third order susceptibility constant. Consequently, the onset of self-focussing [17] and SG is expected to be faster for higher values of $\chi^{(3)}$. For femtosecond timescales, where only the nonresonant electronic mechanism is relevant, the linearly polarized light induces 1.5 times higher $\Delta n$ than circular [18]. Consequently, our gross observations of lower thresholds (≈1.3 times) and higher SG in case of linear polarization as compared to circular are consistent. Similar arguments hold for the case of water and $CCl_4$ [19]. However, it is important to note that this correspondence should be taken only qualitatively as a quantitative theory that relates $\chi^{(3)}$ to the thresholds of SG is not yet clear.

Beyond threshold for SG, the $I_{SG}$ increases smoothly as the incident laser polarization rotates from circular through elliptical to linear at any fixed intensity of the incident laser, independent of the nature of the isotropic sample (Figure 4). As expected, there is no difference in SG between right and left circular polarization. However, it is important to note the flattening in SG, over a small ellipticity range, after the initial rise from circular, independent of the sample and the intensity of the incident laser. This behavior is perhaps a manifestation of changes in mechanisms of SG for different ellipticity ranges and is currently under further investigation.

The suppression of SG in circular polarization can be quantified by defining a ratio R between the linear and circular polarization. We find that this ratio $R=(I_{SG})_{Linear}/(I_{SG})_{Circular}$ of the overall SG is intensity dependent. Beyond thresholds for linear and circular, R decreases from a value of >10, with increasing intensities to reach an asymptotic value of ≈1.6 (Fig. 3b) almost independent of the samples. The large initial values of R (>50) are due to the rise of the SG from linear polarization alone while the threshold for the circular polarization is yet to be reached. In the total supercontinuum generated, we expect contributions from nonlinear processes like Coherent Anti-Stokes Raman and parametric four-photon interactions etc and that have been previously invoked in explaining supercontinuum studies [5,9,10]. Each of these processes would have its own characteristic onset, contribution, and intensity dependence that can usually be modeled with an exponential growth $I_{SG} \sim \exp[G \bullet I_p]$. The gain factor (G) in the exponential is proportional to the respective $\chi^{(3)}$ for each process, and $\chi^{(3)}$ for linear polarization is higher than for circular [18].  Thus, if we invoke the simple $\chi^{(3)}$ dependent exponential gain argument, we expect the ratio R to increase with $I_P$, which is contrary to our results.

Our intensity dependence of SG (figure 3a) shows an initial sharp growth, which gets slower as input laser intensity $I_p$ increases.  Without getting into details of the individual processes, we find that we can fit all our total SG data quite reliably (as shown in Fig. 3a) with an exponential growth model that uses intensity dependent gain.  The functional form for the total supercontinuum generated is: $I_{SG}=I_0 \bullet \exp[G \bullet (I_P-I_{th})]$, where $I_0$ is the background, $I_{th}$ is the threshold for the onset of supercontinuum, and the intensity dependent gain $G = \alpha/(1+\beta(I_P-I_{th}))$, which includes $\alpha$ as the gain constant and (with order of magnitude $\approx 10^{-10}$ cm$^2$/W for 1-cm long samples) and $\beta$ as the gain suppressing parameter (that is about an order of magnitude smaller than $\alpha$).  The higher value of the parameter $\alpha$ for linear as compared to circular is expected as $\chi^{(3)}$ for linear is higher than circular.  However, suppression parameter $\beta$ is even more sensitive to polarization, which implies that the reverse non-linear channels at high energies cause even greater slowing of SG gain in linear than in circular.  Thus, we get the correct suppression behavior where R decreases with $I_P$ (Figure 3b).  Our use of the intensity dependent gain parameter to explain the counterintuitive ratio decrease could perhaps be justified by simultaneous occurrence of processes like Inverse Raman Scattering, or frequency conversion [10] that might decrease the SG gain with intensity.  In other words, at higher intensities, other nonlinear channels open and this depletes energy from SG leading to a slowing of the initial growth rate.

In conclusion, the results presented here show that incident laser polarization is an important parameter that can be used to control the SG process.  This control is important considering the present trend of applications of SG as indicated earlier, either in its more efficient generation or in suppression, as the case may be. The suppression observed in circular polarization could lead to superior data transfer and transport of high-energy laser beams in optical fibers, which would be of immediate relevance in optical communications and beam delivery systems.

We thank M. Krishnamurthy, G. Ravindra Kumar and D. Mathur for many fruitful discussions. We also thank V. Kumarappan for help with Scribble; the software used for data acquisition. The femtosecond laser system used in these experiments was partially funded by Department of Science and Technology, Government of India.

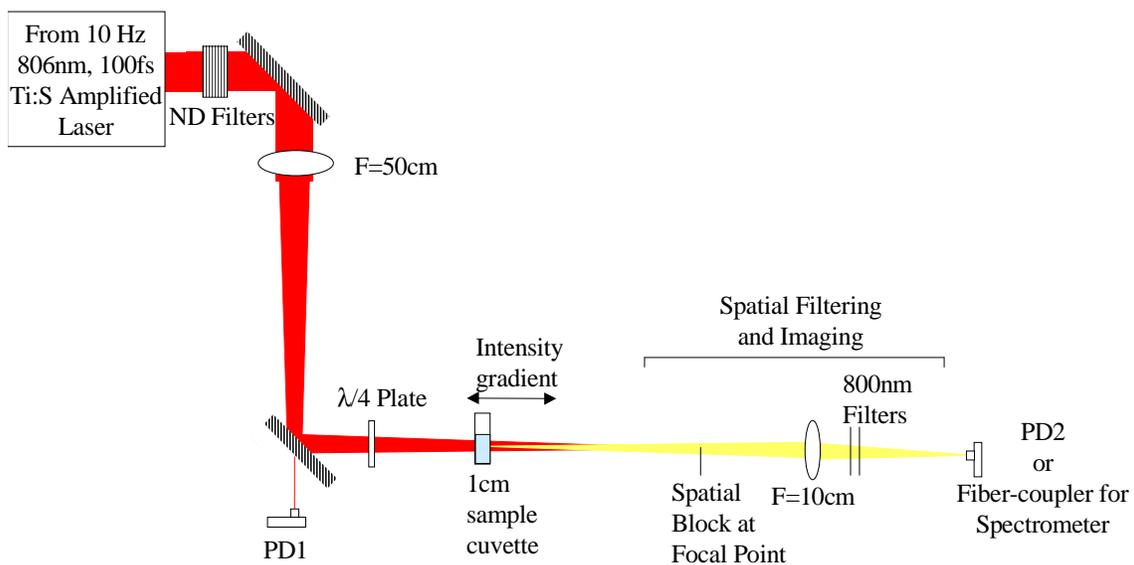

Figure 1. Experimental SG setup showing both the collection techniques: single-shot spectra measurements and a dual photodiode (PD1, PD2) measurement scheme.

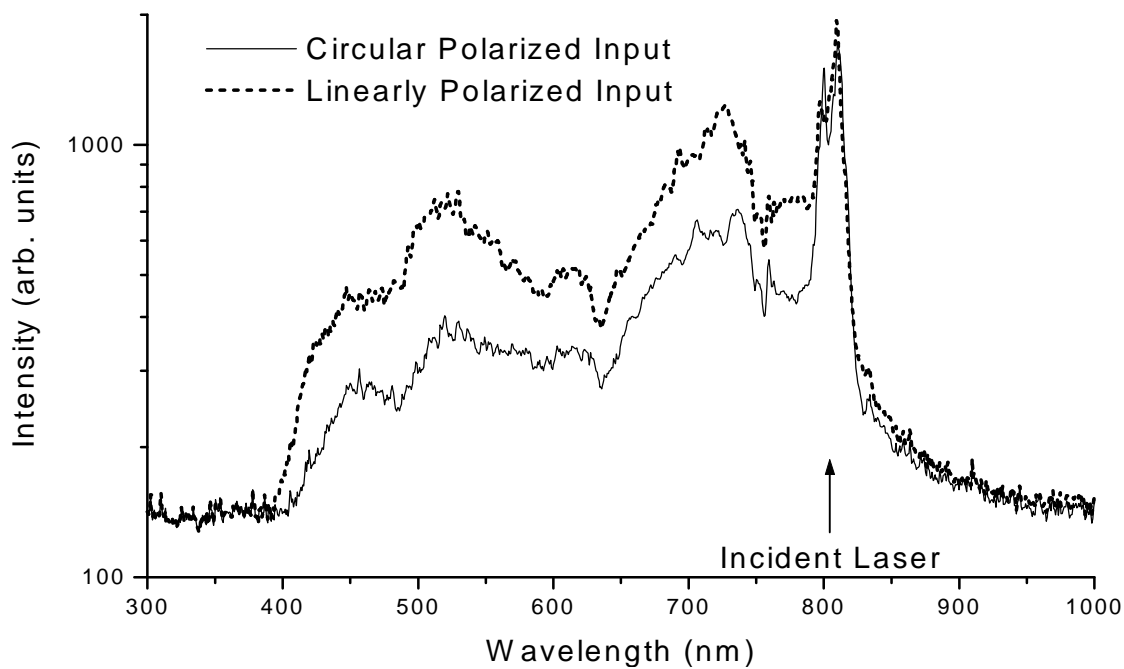

Figure 2. Continuum generated in pure water for linearly versus circularly polarized 100fs Ti:Sapphire laser at identical input intensities. This particular result is near the asymptotic ratio limit at incident laser intensities of $1.1\times10^{11}$ W/cm$^2$.

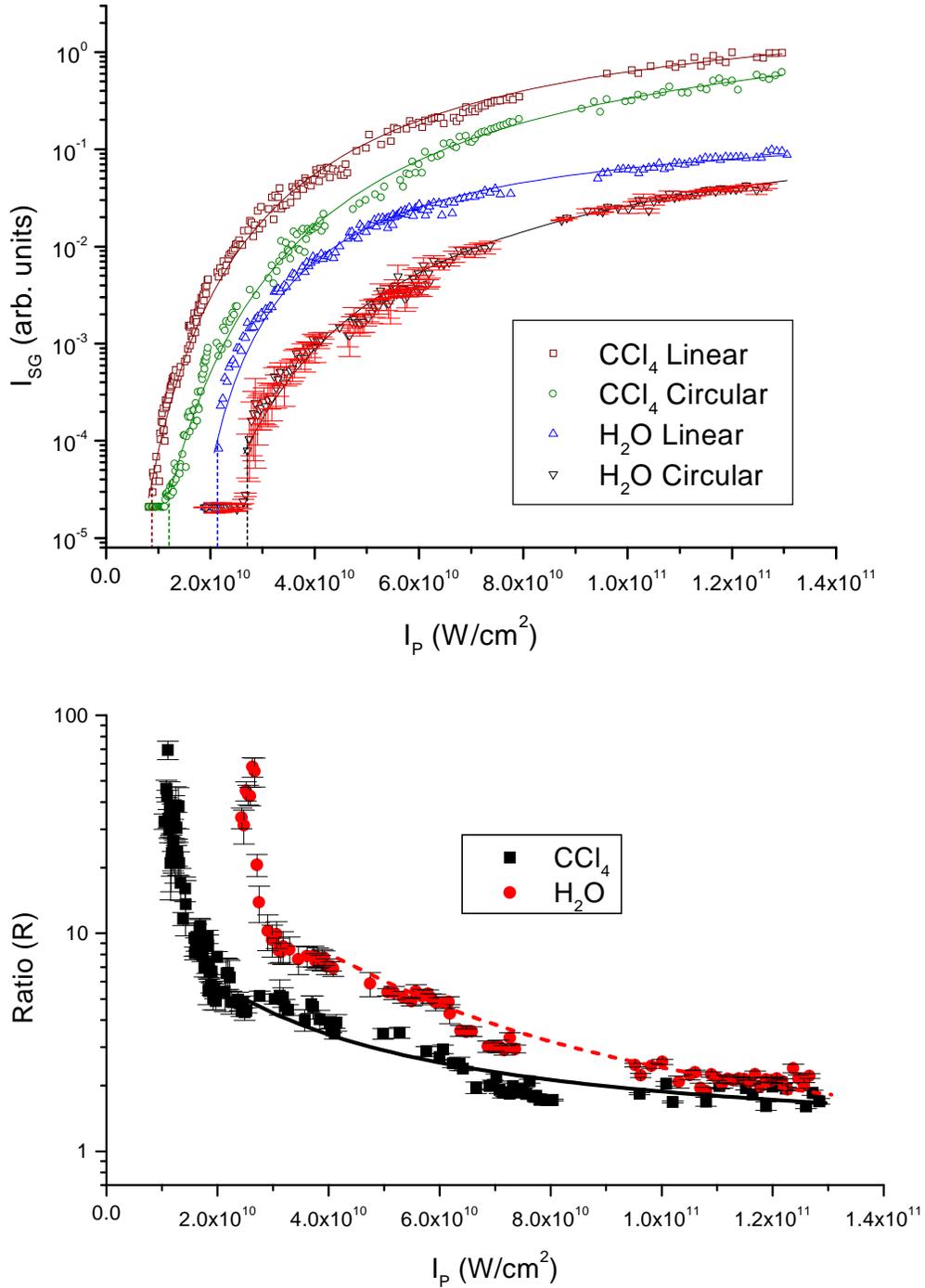

Figure 3. (a) Variation of total SG intensity $I_{SG}$ with input laser intensity $I_P$. The solid curve represents a fit to the actual data points with the model as discussed in the text. The dashed lines show the threshold for the onset of supercontinuum. The error-bars are shown only on one dataset to maintain visual clarity, all other error-bars are similar.
(b) Plot of the ratio of linear to circular as a function of incident laser intensity. The actual error-bars for the data is also shown on the plot. The solid and dashed curves are the ratios obtained based on our model discussed in the text.

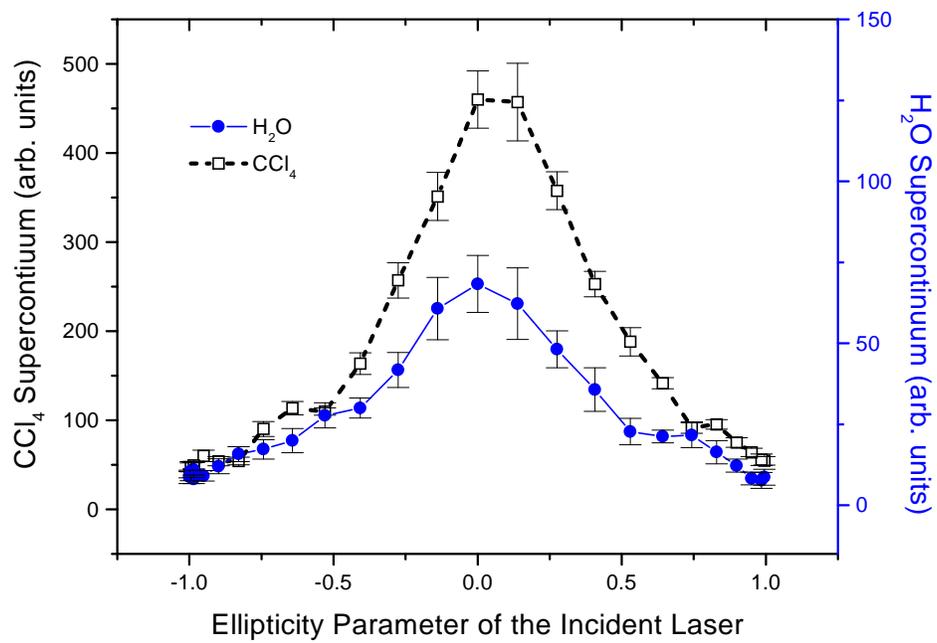

Figure 4. The typical ellipticity dependence of SG in $CCl_4$ and $H_2O$ as a function of the incident laser polarization.